# Multiple radio frequency measurement with an improved frequency resolution based on stimulated Brillouin scattering with a reduced gain bandwidth

Taixia Shi[a,b] and Yang Chen[a,b,*]

[a] *Shanghai Key Laboratory of Multidimensional Information Processing, East China Normal University, Shanghai 200241, China*
[b] *Engineering Center of SHMEC for Space Information and GNSS, East China Normal University, Shanghai 200241, China*
[*] *ychen@ce.ecnu.edu.cn*

**ABSTRACT**
A photonic-assisted multiple radio frequency (RF) measurement approach based on stimulated Brillouin scattering (SBS) and frequency-to-time mapping with high accuracy and high-frequency resolution is reported. A two-tone signal is single-sideband (SSB) modulated on an optical carrier via a dual-parallel Mach-Zehnder modulator to construct one SBS gain and two SBS losses for SBS gain bandwidth reduction. The unknown RF signal is also SSB modulated on a carrier that has been modulated by a sweep signal, thus the unknown RF signal is converted to a sweep optical signal along with the sweep optical carrier. The bandwidth-reduced SBS gain spectrum is detected by the sweep optical signals at different specific time, mapping the RF frequencies to the time domain. An experiment is performed. RF frequencies from 0.3 to 7.6 GHz are simultaneously measured with a root mean square error of less than 1 MHz. In addition, the frequency resolution of the measurement can be much lower than 10 MHz, which is now the best result in the RF frequency measurement methods employing the SBS effect.

## 1. Introduction

Radio frequency (RF) measurements play a very important role in many applications such as electronic warfare and radar [1]. With the rapid development of electronic systems, practical applications require higher and higher measurement bandwidth, measurement accuracy, and measurement resolution to acquire the frequency information of unknown intercepted signals. However, the electrical-based frequency measurement system has the disadvantages of limited operating frequency and bandwidth and susceptibility to electromagnetic interference. The photonic-assisted RF frequency measurement system has many unique advantages, such as high frequency, large bandwidth, low loss, good tunability, and immunity to electromagnetic interference, which can overcome the shortcomings of the electrical-based RF frequency measurement methods [2-3].

Various photonic-assisted RF frequency measurement methods have been proposed, in which the methods based on stimulated Brillouin scattering (SBS) effect are widely utilized [4-10] owing to its narrow bandwidth [11]. The SBS-based methods can acquire the frequency by constructing an amplitude comparison function(ACF) [4-6] or mapping the frequency to the time domain [7-10]. The ACF can be obtained by using the property of single-mode fiber (SMF) dispersion [4] or using the SBS effect to attenuate/amplify two adjacent frequency components [5]. These two methods can only be used for single-frequency measurements and the best measurement accuracy is 8 MHz. In [6], the chip-based method can measure multiple RF signals with a measurement error of 1 MHz. However, the minimum frequency interval is 50 MHz, which is mainly limited by the bandwidth of the SBS gain. The methods via mapping the frequency to the time domain can be achieved by converting the phase modulation to amplitude modulation [7-8], in which the measurement error and resolution are about a few tens of megahertz. The methods of mapping the frequency to time-domain can also be obtained by amplifying the probe wave using the SBS gain generated by the unknown RF signal [9], in which the measurement accuracy and resolution are 5 and 18 MHz, respectively. In [10], we proposed a two-step accuracy improvement method with a measurement accuracy of less than 1 MHz and a measurement resolution of around 20 MHz. As can be seen, the minimum distinguishable frequency difference for all of the reported methods is around 20 MHz, which is mainly limited by the bandwidth of the SBS gain spectrum. Furthermore, when the frequency difference of the unknown signal is approaching the minimum distinguishable frequency difference, the measurement error increases due to the overlap of two SBS gains [10].

In this letter, we propose and experimentally demonstrate a multiple RF frequency measurement approach with a high accuracy and an improved frequency resolution. One dual-parallel Mach-Zehnder modulator (DP-MZM) and two RF signals are used to construct one SBS gain and two SBS losses to reduce the bandwidth of the SBS gain spectrum [11], thus greatly increasing the RF frequency measurement resolution and also the measurement accuracy especially when the frequency difference of the unknown RF signal is very small. A proof-of-concept experiment is performed. RF frequencies from 0.3 to 7.6 GHz are simultaneously measured with a root mean square error (RMSE) of less than 1 MHz. In addition, the frequency resolution of the measurement can be much lower than 10 MHz. To the best of our knowledge, this is the best measurement resolution in the RF frequency measurement methods employing the SBS effect.

## 2. Principle

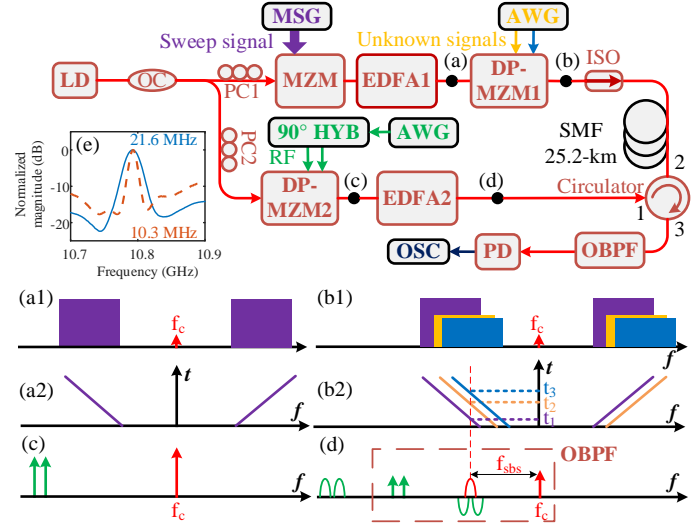

Fig. 1. The schematic diagram of the proposed RF frequency measurement system. LD, laser diode; OC, optical coupler; PC, polarization controller; MZM, Mach-Zehnder modulator; EDFA, erbium-doped fiber amplifier; DP-MZM, dual-parallel Mach-Zehnder modulator; ISO, optical isolator; SMF, single-mode fiber; PD, photodetector; OBPF, optical bandpass filter; MSG, microwave signal generator; AWG, arbitrary waveform generator; 90° HYB, 90° electrical hybrid coupler; OSC, oscilloscope.

The schematic diagram of the proposed system is shown in Fig. 1. An optical carrier centered at 193.08 THz from the laser diode (LD, ID Photonics CoBriteDX1-1-C-H01-FA) is split into two paths through an optical coupler (OC). The optical carrier from the upper branch is sent to a Mach-Zehnder modulator (MZM, Sumitomo T.MXH1.5-20PD-ADC-LV) through a polarization controller (PC1), which is carrier-suppressed double-sideband (CS-DSB) modulated by a sweep signal from a microwave signal generator (MSG, HP, 83752B), as shown in Fig. 1(a). The optical signal from the MZM is amplified by an erbium-doped fiber amplifier (EDFA, MAX-RAY PA-35-B) and then sent to DP-MZM1(Fujitsu 7961EX), where it is single-sideband (SSB) modulated by the unknown RF signals from the arbitrary waveform generator (AWG, Keysight M8195A). Here, two channels of the AWG are used to generate two quadrature signals, where the bias points of DP-MZM1 are tuned slightly away from carrier-suppressed single-sideband (CS-SSB) modulation to generate not only the sidebands but also the carrier itself, as shown in Fig. 1(b). The SSB modulation in DP-MZM1 is not implemented using an MZM because the DP-MZM can achieve an amplitude tunable carrier. The optical signal from DP-MZM1 functions as the probe wave and is injected into a section of 25.2-km SMF through an optical isolator. In the lower branch, the optical signal is injected into DP-MZM2 (Fujitsu 7961EX) through PC2, where it is CS-SSB modulated by a two-tone RF signal from the AWG via an electrical amplifier (EA, Multilink MTC5515) and a 90° electrical hybrid coupler (90° HYB, Macon Omni-Spectra FSC). Same as in DP-MZM1, the carrier is not fully suppressed and is tuned to make its power approximately equal to its sideband, as shown in Fig. 1(c). The output signal from DP-MZM2 is amplified by EDFA2 (Amonics, EDFA-PA-35-B) and launched into the 25.2 km SMF via an optical circulator, where it interacts with the counter-propagating wave from the upper branch. The two optical sidebands and the optical carrier of the pump wave will generate two SBS losses and an SBS gain, as shown in Fig. 1(d). To reduce the bandwidth of the SBS gain spectrum, the frequencies of the two-tone signal must be carefully selected. In the experiment, the frequencies are set to $2 \times f_{sbs} \pm 14.5$ MHz. Under these circumstances, a narrow SBS gain spectrum is constructed by one SBS gain at $f_c - f_{sbs}$ and two SBS losses at $f_c - f_{sbs} \pm 14.5$ MHz, as shown in Fig. 1(d). When the sweep optical signals shown in Fig. 1(b) passes through the SMF, the bandwidth-reduced SBS gain will amplify different sweep optical signals at different specific time. For example, in Fig. 1, a sweep carrier and two sweep optical sidebands will be amplified at $t_1$, $t_2$, and $t_3$, respectively. Therefore, the sweep carrier and sweep optical sidebands will be filtered by the narrowed SBS gain at the time around $t_1$ and $t_2$ respectively, so the frequency-to-time mapping is achieved. Then, the optical signal from the SMF is sent to an optical bandpass filter (OBPF) to filter out most of the amplified spontaneous emission noise of the EDFA. The filtered optical signal is then detected in a photodetector (PD, Discovery Semiconductor DSC-40S) and monitored by an oscilloscope (OSC, Rohde & Schwarz, RTO2000).

## 3. Experimental results and discussion

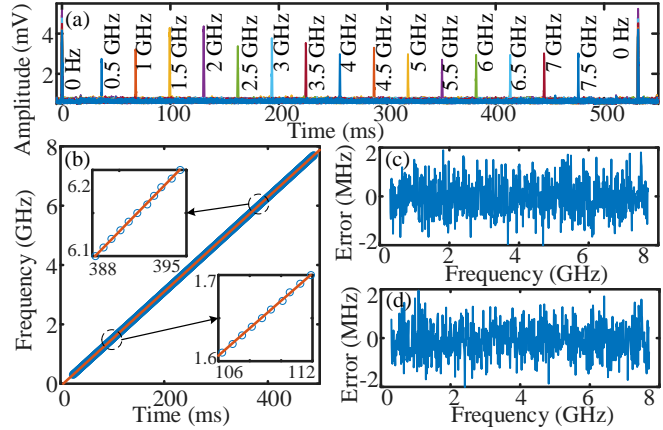

Fig. 2. Single-frequency measurements. (a) The photocurrents at the output of the PD when an RF signal ranging from 0.5 to 7.5 GHz with a frequency step of 0.5 GHz are measured; (b) time-frequency diagram fitted and calibrated by the first measurement; the measurement error of (c) the second, and (d) the third measurement using the time-frequency diagram in (b).

The SBS gain spectrum and Brillouin frequency shift $f_{sbs}$ are first measured using a vector network analyzer (VNA, Agilent 8720ES) when the RF signal is not applied to DP-MZM2, which is shown in the blue solid line in Fig. 1(e). The measured $f_{sbs}$ is about 10.792 GHz, and the 3-dB bandwidth of the SBS gain spectrum is 21.6 MHz. Then, a two-tone signal at 21.570 and 21.599 GHz is applied to DP-MZM2 to reduce the SBS gain bandwidth. As shown in the red dotted line in Fig. 1(e), the 3-dB bandwidth of the SBS gain spectrum is reduced to 10.3 MHz. The measurement performance of the system is studied when the SBS gain bandwidth is reduced. In all the following measurements, a 13-dBm RF signal generated from the MSG, which sweeps from 10.5 to 18.5 GHz with a 500 ms period is applied to the MZM. In fact, because the MSG has a bandwidth point at 11 GHz [10], the sweep RF signal can only be used to measure the RF frequency when it sweeps from 11 to 18.5 GHz, corresponding to a frequency measurement range from 0.208 to 7.708 GHz. In the experiment, the RF signal sweeps from 10.5 GHz to ensure that the sweep carrier from DP-MZM2 can be detected by the SBS gain spectrum, which is further used as a reference to calibrate the time-frequency diagram of the measurement system.

For single frequency measurements, a 7-dBm unknown RF signal generated from the AWG ranging from 0.3 to 7.7 GHz with a frequency step of 0.01 GHz is applied to DP-MZM1. Then, three measurements are recorded by the OSC. The sampling rate of the OSC is set to 100 kSa/s, which is used for all of the following measurements. Fig. 2(a) shows some of the photocurrents from the PD in the first measurement with a frequency step of 0.5 GHz. The first and last pulses corresponding to the sweep carrier are further used as a reference in our measurement, whereas the other pulses corresponding to different sweep optical sidebands are generated from the unknown signal from 0.5 to 7.5 GHz with a frequency step of 0.5 GHz. Fig. 2(a) shows that the time interval between the 0 Hz pulse and the 0.5 GHz pulse is larger than the subsequent pulse intervals also generated by two frequencies with a frequency difference of 0.5 GHz, which is mainly caused by the bandwidth point of the MSG at 11 GHz (a switching time is introduced when passing the bandwidth point). In the experiment, the time differences $t_m$ between the signal pulse and reference pulse corresponds to the frequency $f_m$ of the measured signal. Therefore, a linear fitting curve is obtained from the 741 results from the first single frequency measurement, as shown in Fig. 2(b). The time-frequency relationship can be expressed as

$$f_m = 15.923 \times t_m - 81.390. \tag{1}$$

The linear fitting equation is used for all of the following measurements. Fig. 2(c) and 2(d) show the measurement errors of the second and the third measurements when an unknown RF signal ranging from 0.3 to 7.7 GHz with a frequency step of 0.01 GHz is measured. The measurement errors are all less than ±2.1 MHz, and the root means square errors (RMSEs) of the measurement errors are 0.6886 and 0.6558 MHz, respectively.

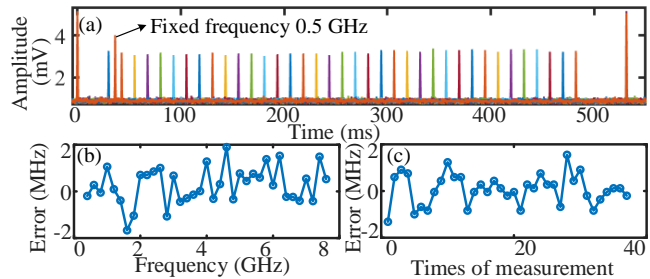

Fig. 3. Dual-frequencies measurements. (a) The photocurrents at the output of the PD when an unknown RF signal fixed at 0.5 GHz and another unknown RF signal ranging from 0.4 to 7.6 GHz with a frequency step of 0.2 GHz are measured. The measurement errors of (b) the unknown RF signal ranging from 0.4 to 7.6 GHz and (c) the fixed unknown RF signal at 0.5 GHz.

Dual-frequency measurements are then studied using the proposed system. A 4-dBm unknown RF signal fixed at 0.5 GHz and another 4-dBm unknown RF signal with its frequency changed from 0.4 to 7.6 GHz with a frequency step of 0.2 GHz generated from AWG are applied to DP-MZM1. Fig. 3(a) shows the photocurrents at the output of the PD. The first and last pulses correspond to the sweep carrier are also used as the reference. The other pulses correspond to the unknown RF signals. As can be seen, the pulses of the fixed unknown RF signal at 0.5 GHz are overlapped. Based on the time-frequency relationship of the first single frequency measurement, Fig. 3(b) and 3(c) show the measurement errors of the unknown RF signal ranging from 0.4 to 7.6 GHz and the unknown fixed RF signal at 0.5 GHz, respectively. The RMSEs of the measurement errors are 0.8174 and 0.6461 MHz, respectively.

The system is then used for four-frequency measurements. Three 1-dBm unknown RF signals fixed at 0.5, 1.5, and 2.5 GHz and another 1-dBm unknown RF signal with its frequency changed from 0.4 to 7.6 GHz with a frequency step of 0.2 GHz are generated by AWG and then applied to DP-MZM1. Fig. 4(a) shows the photocurrents at the output of the PD. Based on the time-frequency relationship of the first single frequency measurement, Fig. 4(b) and (c) show the measurement errors of the unknown RF signal ranging from 0.4 to 7.6 GHz and the unknown fixed RF signals. The measurement errors are all less than ±2.5 MHz. The measurement RMSEs of the fixed unknown RF signal at 0.5, 1.5, and 2.5 GHz and the unknown RF signal ranging from 0.4 to 7.6 GHz are 0.8572, 0.8902, 0.7841, and 0.8116 MHz, respectively. As can be seen from Fig. 4 (b) and (c), the measurement errors have more positive values than negative values. The main reason is the long time interval between the first single frequency measurement and the four frequency measurements in our experiment, during which, the sweep characteristic of the sweep signal from the MSG has changed slightly. Therefore, in practical applications, the deviation of the measurement errors shown in Fig. 4 (b) and (c) can be corrected by another calibration and fitting as done in Fig. 2(b) after a long-time operation.

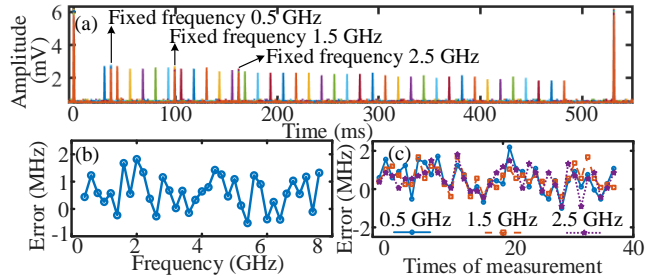

Fig. 4. Four-frequency measurements. (a) The photocurrents at the output of the PD when three unknown RF signals fixed at 0.5 GHz, 1.5 GHz, and 2.5 GHz and another unknown RF signal ranging from 0.2 to 7.6 GHz with a frequency step of 0.2 GHz are measured. The measurement errors of (b) the unknown RF signal ranging from 0.4 to 7.6 GHz and (c) the fixed unknown RF signal at 0.5, 1.5, and 2.5 GHz.

Since an SBS gain and two SBS losses are employed in the system to achieve a much narrower SBS gain spectrum, as shown in Fig. 1 (e), the measurement resolution of the proposed system is much improved compared with all the SBS-based frequency measurement approach. Then, the improvement of measurement resolution is further studied. A 4-dBm RF signal fixed at 0.5 GHz and another 4-dBm RF signal with its frequency changed from 0.45 to 0.55 GHz with a frequency step of 0.01 GHz generated by the AWG are applied to DP-MZM1. Fig. 5(a) and (c) show the measurement results with SBS gain bandwidth reduction. Fig. 5(b) and (d) show the measurement results without SBS gain bandwidth reduction. As can be seen, the pulse width in Fig. 5(a) is narrower than that in Fig. 5(b) because of the SBS gain bandwidth reduction. When the SBS gain bandwidth reduction is disabled, the two adjacent pulses representing two frequencies with 10-

MHz interval can still be roughly distinguished when the frequency interval is 10 MHz, as shown in Fig. 5(d). In comparison, as shown in Fig. 5(c), when SBS gain bandwidth reduction is enabled, the two pulses can still be distinguished very well. Fig. 5(e) and (f) show the measurement errors of the RF signals ranging from 0.45 to 0.55 GHz and fixed at 0.5 GHz. The frequencies are still accurately measured. However, we can clearly know that if the frequency interval is further reduced to less than 10 MHz, it will be difficult to distinguish the adjacent pulses without the SBS gain bandwidth reduction. From Fig. 5 (c), we can expect that the adjacent pulses can still be distinguished using the proposed system with SBS gain bandwidth reduction.

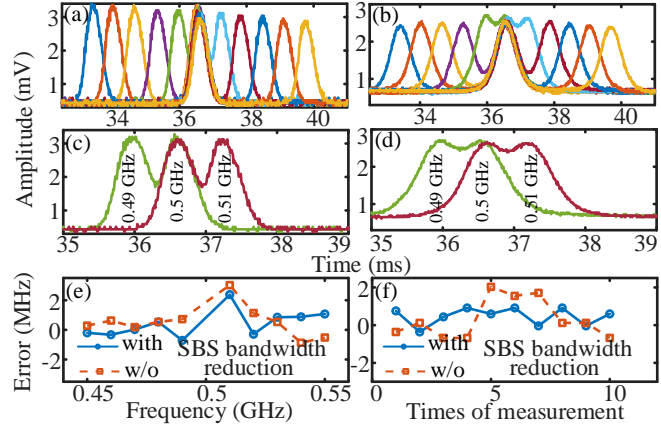

Fig. 5. Improvement of the measurement resolution. The photocurrents from the PD when an RF signal fixed at 0.5 GHz and another RF signal ranging from 0.45 to 0.55 GHz with a frequency step of 0.01 GHz are used, (a) and (c) with SBS gain bandwidth reduction, (b) and (d) without SBS gain bandwidth reduction. The measurement errors of the RF signals (e) ranging from 0.45 to 0.55 GHz and (f) fixed at 0.5 GHz.

The advantages of the proposed method include high measurement resolution and accuracy, multiple RF frequency measurement capability, and low signal power required for measurement.

Due to the bandwidth of the SBS gain spectrum, the methods using the SBS effect to measure the RF frequency are commonly limited in resolution. The frequency measurement resolution of the proposed method is higher than all the other existing methods based on the SBS effect because of the SBS gain bandwidth reduction. The accuracy of the proposed method is close to our recent work in [10], which can be further improved by smoothing the waveform as is done in [10]. Nevertheless, when the waveform smooth is introduced to increase the measurement accuracy, the frequency resolution may be influenced when two frequencies are very close to each other.

In [10], because the optical signal generated by the signals to be measured is used as the pump wave and certain pump power is required to generate the SBS gain, the number of frequencies that can be measured is limited because the limited total pump power and limited number of SBS gains that can be generated. In the proposed system, the pump wave is used to generate a fixed SBS gain and the signals to be measured are used as the probe wave to detect the SBS gain, so the number of frequencies that can be measured by the proposed scheme is much greater than the previous work. Furthermore, because the signals to be measured are used as the probe wave, the power that can be measured by the proposed system is significantly reduced compared with that in our previous work [10].

In the experimental verification, the frequency measurement range is demonstrated from 0.3 to 7.6 GHz due to the bandwidth point of the MSG used in the experiment, which has already been discussed in the part of the experiment. In theory, the frequency measurement range of the system can be further increased and shifted by using a sweep signal source with larger bandwidth and shifting the optical signal in the lower branch as we have done in [10].

## 4. Conclusions

In summary, a multiple RF frequency measurement method with an improved frequency resolution based on SBS effect is proposed and experimentally demonstrated. The key significance of the work is that the reduced SBS gain spectrum is used to increase the RF frequency measurement resolution. An experiment is performed. Multiple RF frequencies from 0.3 to 7.6 GHz are accurately measured with a resolution of less than 10 MHz and an RMSE of the measurement error of less than 1 MHz. To the best of our knowledge, the measurement resolution of less than 10 MHz in this letter is the best results in the RF frequency measurement methods employing the SBS effect. Furthermore, the proposed system is promised to have an extensible measurement range by firstly shifting the optical signal injected into the lower branch.

**Funding**

National Natural Science Foundation of China (NSFC) (61971193); Natural Science Foundation of Shanghai (20ZR1416100); Open Fund of State Key Laboratory of Advanced Optical Communication Systems and Networks, Peking University, China (2020GZKF005).

**Conflicts of interest**

The authors declare no conflicts of interest.

**References**

1. N. Filippo, Introduction to Electronic Defense Systems, 2nd ed. (SciTech, 2006).
2. J. Yao, "Microwave Photonics," J. Light. Technol. 27(3), 314-335 (2009).
3. X. Zou, B. Lu, W. Pan, L. Yan, A. Stöhr, and J. Yao, "Photonics for microwave measurements," Laser Photon. Rev. 10(5), 711-734 (2016).
4. W. Li, N. Zhu, and L. Wang, "Brillouin-assisted microwave frequency measurement with adjustable measurement range and resolution," Opt. Lett. 37(2), 166-168 (2012).
5. D. Wang, L. Pan, Y. Wang, Q. Zhang, C. Du, K. Wang, W. Dong, and X. Zhang, "Instantaneous microwave frequency measurement with high-resolution based on stimulated Brillouin scattering," Opt. Laser Technol. 113, 171-176 (2019).
6. H. Jiang, D. Marpaung, M. Pagani, K. Vu, D.-Y. Choi, S. J. Madden, L. Yan, and B. J. Eggleton, "Wide-range, high-precision multiple microwave frequency measurement using a chip-based photonic Brillouin filter," Optica 3(1), 30-34 (2016).
7. S. Zheng, S. Ge, X. Zhang, H. Chi, and X. Jin, "High-Resolution Multiple Microwave Frequency Measurement Based on Stimulated Brillouin Scattering," IEEE Photon. Technol. Lett. 24(13), 1115-1117 (2012).
8. K. Wu, J. Li, Y. Zhang, W. Dong, X. Zhang, and W. Chen, "Multiple microwave frequencies measurement based on stimulated Brillouin scattering with ultra-wide range," Optik 126, 1935-1940 (2015).
9. W. Jiao, K. You, and J. Sun, "Multiple Microwave Frequency Measurement With Improved Resolution Based on Stimulated Brillouin Scattering and Nonlinear Fitting," IEEE Photon. J. 11(2), 5500912 (2019).

10. J. Liu, T. Shi, and Y. Chen, "High-Accuracy Multiple Microwave Frequency Measurement With Two-Step Accuracy Improvement Based on Stimulated Brillouin Scattering and Frequency-to-Time Mapping," J. Light. Technol. 39(7), 2023-2032 (2021).
11. S. Preussler and T. Schneider, "Stimulated Brillouin scattering gain bandwidth reduction and applications in microwave photonics and optical signal processing," Opt. Eng. 55(3), 031110 (2015).